\documentstyle[12pt,amsfonts]{article}
\parskip=\medskipamount

\def\appendix#1{
  \addtocounter{section}{1}
  \setcounter{equation}{0}
  \renewcommand{\thesection}{\Alph{section}}
 \section*{Appendix \thesection\protect\indent
 \parbox[t]{11.715cm} {#1}}
 \addcontentsline{toc}{section}{Appendix \thesection\ \ \ #1}
  }

\renewcommand{\thefootnote}{\fnsymbol{footnote}}

\newcommand {\cD}{{\cal D}}

\newcommand {\cF}{{\cal F}}
\newcommand {\cG}{{\cal G}}

\newcommand {\cL}{{\cal L}}

\newcommand {\cN}{{\cal N}}

\newcommand {\cW}{{\cal W}}

\def\a{\alpha}
\def \bi{\bibitem}

\def\b{\beta}

\def\d{\delta}

\def\G{\Gamma}

\def\m{\mu}

\def\o{\omega}

\def\q{\theta}
\def\r{\rho}
\def\s{\sigma}
\def\t{\tau}

\def\z{\zeta}
\def\D{\Delta}
\def\F{\Phi}
\def\J{\Psi}
\def\L{\Lambda}
\def\O{\Omega}
\def\P{\Pi}

\newcommand{\ad}{{\dot{\alpha}}}                           
\newcommand{\bd}{{\dot{\beta}}}                            
\newcommand{\ve}{\varepsilon}                            

\newcommand{\hf}{\frac12}

\newcommand{\sect}[1]{\setcounter{equation}{0}\section{#1}}

\newcommand{\be}{\begin{equation}}
\newcommand{\ee}{\end{equation}}
\newcommand{\bea}{\begin{eqnarray}}
\newcommand{\eea}{\end{eqnarray}}
\newcommand{\non}{\nonumber}
\begin{document}

\begin{titlepage}
\thispagestyle{empty}

\begin{flushright}
hep-th/0101127 \\
January, 2001
\end{flushright}
\vspace{5mm}

\begin{center}
{\Large\bf  Effective action of \mbox{$\cN = 4$} super Yang-Mills:\\
\mbox{$\cN = 2$}  superspace approach}
\end{center}
\vspace{3mm}

\begin{center}
{\large
S.M. Kuzenko and I.N. McArthur
}\\
\vspace{2mm}

${}$\footnotesize{
{\it
Department of Physics, The University of Western Australia\\
Crawley, W.A. 6009. Australia}
} \\
{\tt  kuzenko@physics.uwa.edu.au},~
{\tt mcarthur@physics.uwa.edu.au}
\vspace{2mm}

\end{center}
\vspace{5mm}

\begin{abstract}
\baselineskip=14pt
Using the $\cN=2$ off-shell formulation
in harmonic superspace for  $\cN=4$ super
Yang-Mills theory, we present a representation
of the one-loop effective action which is free
of so-called coinciding harmonic singularities and
admits a straightforward evaluation of low-energy
quantum corrections in the framework of an $\cN=2$
superfield heat kernel technique.
We illustrate our approach by computing the low-energy
effective action on the Coulomb branch of $SU(2)$
$\cN=4$ super Yang-Mills.
Our work provides the first derivation of the low-energy
action of $\cN=4$ super Yang-Mills theory directly
in $\cN=2$ superspace without
any reduction to $\cN=1$ superfields
and for a generic background
$\cN=2$ Yang-Mills multiplet.
\end{abstract}
\vfill
\end{titlepage}

\newpage
\setcounter{page}{1}

\renewcommand{\thefootnote}{\arabic{footnote}}
\setcounter{footnote}{0}
\sect{Introduction}

Harmonic superspace  \cite{GIKOS,Z}
is a manifestly supersymmetric construction
which allows the description of general $\cN=2$ super
Yang-Mills theories in terms of unconstrained superfields
whilst preserving the $SU(2)_R$ automorphism symmetry
of the $\cN=2$ Poincar\'e superalgebra. The usual superspace variables
are supplemented by harmonics on an internal $S^2.$ The Feynman rules
in this approach were elaborated early in the history of harmonic
superspace \cite{GIOS}.
Less well-developed  is the background field formulation
in harmonic superspace \cite{BBKO,BK2}, which is expected
to be vital for computing
low-energy effective actions in quantum $\cN=2$ super
Yang-Mills theories (the background field formulation
of \cite{BBKO,BK2} has already been used to prove
the $\cN=2$ non-renormalization theorem \cite{BKO}
and to compute the leading non-holomorphic quantum corrections
in the $\cN=4$ super Yang-Mills theory \cite{BK2,BBK};
see also \cite{BP}).

Until now, applications of  the background field formulation
in harmonic superspace have been incomplete at the one-loop level
because of a technical
problem. This problem is characteristic of all
$\cN=2$ super Yang-Mills theories, but it manifests
itself most explicitly in the case of $\cN=4$ SYM,
which is a special $\cN=2$ model in which no
holomorphic quantum corrections are present.
The one-loop effective action of $\cN=4$ SYM consists
of two contributions, each of which contains harmonic distributions
on the internal $S^2$
at coinciding points, and hence singularities \cite{BK2}.
Such singularities of harmonic supergraphs
were first discussed in \cite{GNS}.
Unlike ultraviolet divergences in quantum field theory,
the coinciding harmonic singularities have no physical origin;
they can appear only at intermediate stages of calculations
and must cancel each other in the final expressions
for physical observables. A precise mechanism for
the cancellation of coinciding harmonic singularities
was found in \cite{BK2} only for the case when the background
$\cN=2$ gauge multiplet satisfies the
equations of motion. The use of such a background was sufficient
for the authors of \cite{BK2,BBK}
to compute the leading non-holomorphic quantum correction
$H (W,\bar W )$ on the Coulomb branch of $\cN=4$ SYM
(which was also derived in \cite{gr,gkpr,lvu}
by different methods).

In the present paper, we present a resolution of
the problem of coinciding harmonic singularities in the case
of $\cN=4$ SYM and obtain a representation for the one-loop
effective action which is  suited to  computation
of the low-energy effective action in the
framework of an $\cN=2$ superfield heat kernel technique
(similar to the standard $\cN=1$ results \cite{BK,MG,PB,BKT}). Full
details will appear in a following paper \cite{KM}.
Our work provides the first derivation of the low-energy
action of $\cN=4$ SYM  directly
in $\cN=2$ superspace without
any reduction to $\cN=1$ superfields
and for a generic background
$\cN=2$ Yang-Mills multiplet.

Our $\cN=2$ harmonic superspace conventions correspond
to those adopted in ref.
\cite{GIOS}, apart from the fact that the Greek
letter `$\z$' is used to denote all the coordinates of the analytic
subspace, with ${\rm d}\z^{(-4)}$ the analytic subspace measure.
The algebra
of $\cN=2$ gauge covariant derivatives
$\cD_A = (\cD_a, \cD^i_\a , {\rm \cD}^\ad_j)$
was established in \cite{GSW};
the basic anti-commutation relations can be expressed
\bea
&
\{ \cD^+_\a ,  \cD^+_\b \}
= \{ {\bar \cD}^+_\ad , {\bar \cD}^+_\bd \}
=\{ \cD^+_\a , {\bar \cD}^+_\bd \} = 0~, \non \\
& \{ {\bar \cD}^+_\ad , \cD^-_\a \}
= - \{ \cD^+_\a , {\bar \cD}^-_\ad \}
=2{\rm i}\, \cD_{\a \ad}~, \non \\
&\{ \cD^+_\a ,  \cD^-_\b \}
= -  2{\rm i}\,\ve_{\a \b} {\bar \cW}~, \qquad
\{ {\bar \cD}^+_\ad , {\bar \cD}^-_\bd \}
= 2{\rm i}\,\ve_{\ad \bd} \cW~,
\eea
where $\cD^\pm_\a = \cD^i_\a \,u^\pm_i$,
${\bar \cD}^\pm_\ad = {\bar \cD}^i_\ad \,u^\pm_i$,
with $u^+_i$ and $u^-_i$ the harmonics \cite{GIKOS}.
The field strength $\cW$ satisfies the Bianchi
identities \cite{GSW}
\be
{\bar \cD}^i_\ad \, \cW=0~, \qquad
\cD^{\a (i}\cD_\a^{j)}\, \cW =
{\bar \cD}_{\ad}^{(i} {\bar \cD}^{j) \ad} \,
{\bar \cW} =0~.
\ee
In harmonic superspace, a full set of gauge covariant
derivatives includes the
spherical derivatives
$(\cD^{++}, \, \cD^{--},\,  \cD^0),$ which form
the algebra $su(2)$ and satisfy special
commutation relations with $\cD^\pm_\a$ and
${\bar \cD}^\pm_\ad$ \cite{GIOS}.
Here, $\cD^0$ is the harmonic $U(1)$ charge operator,
and the important property of $\cD^{++}$ is that it
preserves analyticity.
Throughout this paper, we use the so-called
$\t$-frame \cite{GIKOS,GIOS}, and the adjoint
representation of the gauge group is assumed.

\sect{The one-loop effective action}

A formal definition of the one-loop effective action
for $\cN=4$ super Yang-Mills theory has been given
in ref. \cite{BK2}:
\be
\G^{(1)} =
\frac{{\rm i}}{2}\,{\rm Tr}\,{}_{(2,2)} \, \ln
{\stackrel{\frown} {\Box}}{}
- \frac{{\rm i}}{2}\,{\rm Tr}\,{}_{(0,4)} \, \ln
{\stackrel{\frown}{\Box}}{} ~,
\label{4}
\ee
where ${\stackrel{\frown} {\Box}}$ is the analytic d'Alembertian
\bea
{\stackrel{\frown}{\Box}}{}&=&
{\cal D}^m{\cal D}_m+
\frac{{\rm i}}{2}({\cal D}^{+\a}\cW){\cal D}^-_\a+\frac{{\rm i}}{2}
({\bar{\cal D}}^+_{\dot\alpha}{\bar \cW}){\bar{\cal D}}^{-{\dot\alpha}}-
\frac{{\rm i}}{4}({\cal D}^{+\a} {\cal D}^+_\a \cW) \cD^{--}\non \\
&{}& +\frac{{\rm i}}{8}[{\cal D}^{+\alpha},{\cal D}^-_\alpha] \cW
+ \frac{1}{2}\{{\bar \cW},\cW \}
\label{2}
\eea
possessing the following important properties
\be
[\cD^+_\a \,,\, {\stackrel{\frown}{\Box}}{}] = 0~, \qquad
[{\bar \cD}^+_\ad \,,\, {\stackrel{\frown}{\Box}}{}] = 0~.
\ee
The traces are functional traces of operators acting on  analytic
superfields
of appropriate $U(1)$ charge. Specifically,
if $\cF^{(p,4-p)} (\z_1, \z_2)$ is the kernel of
an operator acting on the space of covariantly analytic
superfields of $U(1)$ charge $p$,
then
$$
{\rm Tr}\; \cF^{(p,4-p)} = {\rm tr}
\int{\rm d}\z^{(-4)}\, \cF^{(p,4-p)} (\z, \z)~,
$$
where the  trace `tr' is over group indices.
The two contributions to $\G^{(1)}$ can be represented
by path integrals over unconstrained analytic superfields
$u^{++},\; v^{++}$
and $\r^{(+4)}, \; \s$ as follows:
\bea
\left( {\rm Det}_{(2,2)} \,
{\stackrel{\frown}{\Box}}{} \right)^{-1} &=&
\int [{\rm d}u^{++}]\, [ {\rm d} v^{++}]\,
\exp \left\{  {\rm tr} \int  {\rm d}\zeta^{(-4)} \,
u^{++} {\stackrel{\frown}{\Box}}{} \,v^{++} \right\}~, \label{3} \\
\left( {\rm Det}_{(0,4)} \,
{\stackrel{\frown}{\Box}}{} \right)^{-1} &=&
\int [ {\rm d}\r^{(+4)}]\,[ {\rm d} \s]
\exp \left\{   {\rm tr} \int {\rm d}\zeta^{(-4)} \,
\r^{(+4)} {\stackrel{\frown}{\Box}}{} \,\s \right\}~.
\label{3a}
\eea

Both terms in (\ref{4}) contain harmonic singularities due to the
coincidence limit involved in the functional traces.
However, these cancel each other,
and we claim that the resulting  well-defined
expression for the one-loop effective action is
\be
\G^{(1)} = -\frac{\rm i}{2} \int_{0}^{\infty}
\frac{ {\rm d}t}{t}\; {\rm Tr}\; \left(
{\rm e}^{-{\rm i} \,t\,{\stackrel{\frown}{\Box}} }\;
\P^{(2,2)}_{\rm T} \right)~.
\label{starting}
\ee
Here, $\P^{(2,2)}_{\rm T}(\z_1 , \z_2)$ is a projector
on the space of covariantly analytic transverse superfields
$\cG^{++}(\z)$
defined by the constraints
\be
\cD^{+ \hat{\a}} \cG^{++}
= 0~,
\qquad \cD^{++} \cG^{++} = 0~,
\ee
where $\cD^{+ \hat{\a}} = ( \cD^{+ \a},
{\bar \cD}^{+ \ad})$. The  properties of
$\P^{(2,2)}_{\rm T}(\z_1 , \z_2)$ are:
\bea
 \cD^{+ \hat{\a}}_1 \P^{(2,2)}_{\rm T}(\z_1 , \z_2)
&=&  \cD^{+ \hat{\a}}_2 \P^{(2,2)}_{\rm T}(\z_1 , \z_2) =0~, \label{prop1} \\
  \cD^{++}_1  \P^{(2,2)}_{\rm T}(\z_1 , \z_2)
&=& \cD^{++}_2  \P^{(2,2)}_{\rm T}(\z_1 , \z_2)=0~, \label{prop2}\\
 \int {\rm d} \z^{(-4)}_3\,
\P^{(2,2)}_{\rm T}(\z_1 , \z_3) \, \P^{(2,2)}_{\rm T}(\z_3 , \z_2)
&=& \P^{(2,2)}_{\rm T}(\z_1 , \z_2)~, \label{prop3}\\
\Big(\P^{(2,2)}_{\rm T}(\z_1 , \z_2)\Big)^{\rm T} &=&
\P^{(2,2)}_{\rm T}(\z_2 , \z_1)~.\label{prop4}
\eea

The projector $\P^{(2,2)}_{\rm T}$ is related to
the Green's function $G^{(0,0)} (\z_1 ,\z_2) $
for a $\o$--hypermultiplet coupled to background $\cN=2$ gauge
superfields, which satisfies the equation
\be
(\cD_1^{++})^2 G^{(0,0)}(\z_1 ,\z_2) = - \d_A^{(4,0)}(\z_1 ,\z_2)
\label{o-gf}.
\ee
This Green's function can be expressed explicitly
\cite{GIOS,BKO} in the form\footnote{Here
$(\cD^+)^4 = \frac{1}{16} \cD^{+\a} \cD^+_\a \,
{\bar \cD}^+_\ad {\bar \cD}^{+\ad}$.}
\be
G^{(0,0)}(\z_1, \z_2)
=  \frac{1}{{\stackrel{\frown}{\Box}}{}_1}
(\cD_1^+)^4 \,(\cD_2^+)^4
\left\{ \delta^{12}(z_1-z_2)
{(u^-_1 u^-_2)\over (u^+_1 u^+_2)^3}
\right\}~,
\label{8}
\ee
and is manifestly analytic
in both arguments. Less obvious is the fact that
$G^{(0,0)} (\z_1 ,\z_2) $ is also symmetric,
\be
\Big(G^{(0,0)}(\z_1 ,\z_2)\Big)^{\rm T} =
G^{(0,0)}(\z_2 ,\z_1)~.
\ee
The latter property will be discussed
in detail  in \cite{KM}.
It follows from eq. (\ref{o-gf}) that the analytic
two-point function
\be
\P^{(2,2)}_{\rm L}(\z_1 ,\z_2)
=\cD^{++}_1 \cD^{++}_2 G^{(0,0)} (\z_1 ,\z_2)
\ee
has the properties
\bea
 \int {\rm d} \z^{(-4)}_3\,
\P^{(2,2)}_{\rm L}(\z_1 ,\z_3) \, \P^{(2,2)}_{\rm L}(\z_3 ,\z_2)
&=& \P^{(2,2)}_{\rm L}(\z_1 ,\z_2)~, \\
\cD^{++}_1 \P^{(2,2)}_{\rm L}(\z_1 ,\z_2)
&=& \cD^{++}_1 \d_A^{(2,2)}(\z_1 ,\z_2)~.
\eea
Therefore, $\P^{(2,2)}_{\rm L}$ is the projector on the space
of longitudinal analytic superfields
\be
\cF^{++} = \cD^{++} \L~, \qquad
\cD^{+ \hat{\a}} \L = 0~.
\ee
As a result,  $\P^{(2,2)}_{\rm T}$ can be expressed as
\be
\P^{(2,2)}_{\rm T} (\z_1 ,\z_2)
= \d_A^{(2,2)}(\z_1 ,\z_2) ~-~ \P^{(2,2)}_{\rm L}(\z_1 ,\z_2)~,
\label{pt-def}
\ee
which establishes the connection with the Green's function
$G^{(0,0)} (\z_1 ,\z_2). $ When the $\cN=2$ vector multiplet
is set to zero, $\P^{(2,2)}_{\rm T}$ reduces to the flat
projector derived in \cite{GIOS}.

The representation (\ref{starting}) for the one-loop effective action
can  readily be deduced from the formal one
given by eq. (\ref{4}) in the case when the background
gauge superfield satisfies the classical equation of
motion
\be
\cD^{\a (i} \cD^{j)}_\a  \cW = 0 \qquad \Longleftrightarrow \qquad
[\cD^{++}, {\stackrel{\frown}{\Box}}] =0~.
\label{5}
\ee
In this case, one can argue as follows:
\bea
\G^{(1)} &=& -\frac{\rm i}{2} \int_{0}^{\infty}
\frac{ {\rm d}t}{t}\; {\rm Tr} \; \left\{
{\rm e}^{-{\rm i} \,t\,{\stackrel{\frown}{\Box}}_1 }\;
\d_A^{(2,2)}(\z_1,\z_2)
-{\rm e}^{-{\rm i} \,t\,{\stackrel{\frown}{\Box}}_1 }\;
\d_A^{(0,4)}(\z_1,\z_2)
\right\} \non \\
 &=& -\frac{\rm i}{2} \int_{0}^{\infty}
\frac{ {\rm d}t}{t}\; {\rm Tr}\; \left\{
{\rm e}^{-{\rm i} \,t\,{\stackrel{\frown}{\Box}}_1 }\;
\d_A^{(2,2)}(\z_1,\z_2)
+ {\rm e}^{-{\rm i} \,t\,{\stackrel{\frown}{\Box}}_1 }\;
(\cD^{++}_2)^2 G^{(0,0)}(\z_1,\z_2 )\right\} \non \\
&=& -\frac{\rm i}{2} \int_{0}^{\infty}
\frac{ {\rm d}t}{t}\; {\rm Tr} \; \left\{
{\rm e}^{-{\rm i} \,t\,{\stackrel{\frown}{\Box}}_1 }\;
\d_A
^{(2,2)}(\z_1,\z_2)
-\cD^{++}_1 {\rm e}^{-{\rm i} \,t\,{\stackrel{\frown}{\Box}}_1 }\;
\cD^{++}_2 G^{(0,0)}(\z_1,\z_2 )\right\} \non \\
&=& -\frac{\rm i}{2} \int_{0}^{\infty}
\frac{ {\rm d}t}{t}\; {\rm Tr}\; \left\{
{\rm e}^{-{\rm i} \,t\,{\stackrel{\frown}{\Box}}_1 }
\left( \d_A^{(2,2)}(\z_1,\z_2) -\P^{(2,2)}_{\rm L}(\z_1,\z_2) \right)
\right\}~.
\label{line}
\eea
The cyclic property of the functional trace has been used in going
from the second line to the third line.

In the off-shell case, the operators $\cD^{++}$ and
${\stackrel{\frown}{\Box}}$ do not commute
in general\footnote{However, from eq. (\ref{com})
one derives
$[\cD^{++}, {\stackrel{\frown}{\Box}}]\; q^+ = 0$
and  $[(\cD^{++})^2, {\stackrel{\frown}{\Box}}] \;\o = 0$. }:
\be
[\cD^{++}, {\stackrel{\frown}{\Box}}] \F^{(q)} =
\frac{{\rm i}}{4} \,(1-q)(\cD^{+ } \cD^+ \cW) \,  \F^{(q)}
\label{com}
\ee
for an arbitrary analytic superfield $\F^{(q)}$ with $U(1)$ charge $q$.
Thus, the process of  interchanging the operators $\cD^{++}_1$ and
${\rm exp}(-{\rm i}\,t \,{\stackrel{\frown}{\Box}}_1)$
in going from the third line of (\ref{line})
to the fourth line introduces
an extra contribution to the effective action
containing factors of the classical equation of motion.
As is well known, contributions to the effective
action containing factors of the classical equations
of motion are ambiguous and may be ignored.
On the other hand, this extra contribution proves
to involve coinciding harmonic singularities which
must be regularized; for a regularization preserving
analyticity, the whole extra contribution
can be shown to vanish.

It follows from (\ref{prop2}) that $\P^{(2,2)}_{\rm T}$
has the following dependence on the harmonics:
\be
\P^{(2,2)}_{\rm T} (z_1,u_1\,,\,z_2,u_2)=
\P^{(ij)(kl)}_{\rm T}(z_1\,,\,z_2) \;(u^+_1)_i(u^+_1)_j\,
(u^+_2)_k(u^+_2)_l~.
\label{pt-fe}
\ee
Therefore the effective action (\ref{starting}) is
by construction free of harmonic singularities.
We will present a representation for $\P^{(2,2)}_{\rm T}$
which is useful for heat kernel calculations.
Using the standard properties of harmonic distributions
\cite{GIOS} and the identity
\be
[\cD^{++}, \frac{1}{ {\stackrel{\frown}{\Box}}}]
= -\frac{1}{ {\stackrel{\frown}{\Box}}}\;
[\cD^{++},  {\stackrel{\frown}{\Box}}]  \;
\frac{1}{ {\stackrel{\frown}{\Box}}}  ~,
\ee
we deduce from the definition (\ref{pt-def}) of $\P^{(2,2)}_{\rm T}$
that
\bea
\P^{(2,2)}_{\rm T} (\z_1,\z_2) &=&
\frac{\rm i}{4}\;
\frac{1}{ {\stackrel{\frown}{\Box}}_1} \;
(\cD^+_1 \cD^+_1 \cW_1)\;
\frac{1}{ {\stackrel{\frown}{\Box}}_1} \;
(\cD^+_1)^4 (\cD^+_2)^4 \;\d^{12} (z_1 -z_2)\;
\frac{(u^-_1 u^+_2) }{(u^+_1 u^+_2)^{3}} \non \\
&-& \frac{1}{ {\stackrel{\frown}{\Box}}_1} \;
(\cD^+_1)^4 (\cD^+_2)^4 \;\d^{12} (z_1 -z_2)\;
\frac{1 }{(u^+_1 u^+_2)^2}~.
\eea
The next step is to realize that the covariant derivatives $\cD^{+ \hat{\a}}_2$
in a two-point function of the form
\be
(\cD^+_1)^4 (\cD^+_2)^4 \;\d^{12} (z_1 -z_2)\;
\frac{1}{(u^+_1 u^+_2)^q}~,
\ee
can be re-expressed in terms of the covariant derivatives
 $\cD^{\pm \hat{\a}}_1$  with the help of the identity
\be
\J^+_2 = (u^+_1 u^+_2) \, \J^-_1   - (u^-_1 u^+_2) \, \J^+_1~, \qquad \quad
\J^\pm = \J^i \,u^\pm_i
\ee
along with the algebra of gauge covariant derivatives
which implies, in particular, that
\be
(\cD^+)^4 \; \cD^+_{\hat{\a}} = \cD^+_{\hat{\a}}\;(\cD^+)^4 =0~.
\ee
The result is
\bea
 (\cD^+_1)^4 (\cD^+_2)^4 \;
\frac{\d^{12} (z_1 -z_2)  }{(u^+_1 u^+_2)^q} &=&
(\cD^+_1)^4 \;
\Bigg\{ (\cD^-_1)^4 \;
\frac{1 }{(u^+_1 u^+_2)^{q-4}}
- \frac{\rm i}{2} \,
\D^{--}_1\;
\frac{(u^-_1 u^+_2) }{(u^+_1 u^+_2)^{q-3}} \non \\
- {\stackrel{\frown}{\Box}}_1 \;
\frac{(u^-_1 u^+_2)^2 }{(u^+_1 u^+_2)^{q-2}}
&+& \frac{\rm i}{4}(q-3)\;  (\cD^+_1 \cD^+_1 \cW_1)\;
\frac{(u^-_1 u^+_2)^3 }{(u^+_1 u^+_2)^{q-1}} \Bigg\}\;
\d^{12} (z_1 -z_2) ~,
\label{master}
\eea
where
\bea
\D^{--} =\cD^{\a \ad} \cD^-_{\a} {\bar \cD}^-_{\ad}
&+& \hf \cW (\cD^-)^2 + \hf {\bar \cW} ({\bar \cD}^-)^2 \non \\
&+& (\cD^- \cW) \cD^- + ({\bar \cD}^- {\bar \cW}) {\bar \cD}^-
+\hf (\cD^- \cD^- \cW)~.
\eea
It is worth noting that eq. (\ref{master})
simplifies in the two special cases
$q=2$ and $q=3,$ which are of  importance here.
Putting all these results together, we obtain the following
representation for $\P^{(2,2)}_{\rm T}:$
\bea
\P^{(2,2)}_{\rm T}(\z_1, \z_2) &=&
(\cD^+_1)^4 \;\d^{12} (z_1 -z_2)\;(u^-_1 u^+_2)^2 \non \\
&-& \frac{1}{ {\stackrel{\frown}{\Box}}_1}
\left\{ (u^+_1 u^+_2)
- \frac{\rm i}{4}\;  (\cD^+_1 \cD^+_1 \cW_1)\;
\frac{1}{ {\stackrel{\frown}{\Box}}_1} \;
(u^-_1 u^+_2) \right\}\;
\non \\
&\times &   (\cD^+_1)^4
\left\{ (u^+_1 u^+_2)\;(\cD^-_1)^4
-\frac{\rm i}{2} (u^-_1 u^+_2) \;\D^{--}_1 \right\}
\;\d^{12} (z_1 -z_2)~.
\label{working}
\eea
The factors of $1/{\stackrel{\frown}{\Box}}$
are to be understood in the sense of Schwinger's proper-time
representation,
\be
\frac{1}{ {\stackrel{\frown}{\Box}}}= {\rm i}
\int_{0}^{\infty}{\rm d}s\;
{\rm e}^{-{\rm i}\,s\,{\stackrel{\frown}{\Box}}}~.
\ee
It is also important to note that the covariant
transverse projector  $\P^{(2,2)}_{\rm T}$
has the property
\be
{\rm Tr} \;\P^{(2,2)}_{\rm T}~=~0~.
\ee

Using this expression for $\P^{(2,2)}_{\rm
T},$ the following key features  of the representation
(\ref{starting}) for the one-loop effective action are apparent:

${\bf 1.}$ Eq. (\ref{starting}) is free of
coinciding harmonic singularities;

${\bf 2.}$ The representation (\ref{starting})
provides a simple and powerful scheme
for computing the effective action in the framework
of an $\cN=2$ superfield proper-time technique;

${\bf 3.}$ Unlike the approach of ref. \cite{BK2},
the representation (\ref{starting}) is valid for arbitrary
background $\cN=2$ gauge superfields.

Using eq. (\ref{prop2}) and rewriting (\ref{starting}) in the form
\be
\G^{(1)} = -\frac{\rm i}{2} \int_{0}^{\infty}
\frac{ {\rm d}t}{t}\; {\rm Tr}\; \left( \P^{(2,2)}_{\rm T}\;
{\rm e}^{-{\rm i} \,t\,{\stackrel{\frown}{\Box}} }\;
\P^{(2,2)}_{\rm T} \right)~,
\ee
it can be seen that the one-loop effective action takes the form
\be
\G^{(1)} =  \int {\rm d} \z^{(-4)}\;
\cL^{(+4)}_{\rm eff}  [\cW, \bar \cW ]~, \qquad
\cD^+_{\hat{\a}} \cL^{(+4)}_{\rm eff}
= \cD^{++} \cL^{(+4)}_{\rm eff}  =0~,
\ee
with $ \cL^{(+4)}_{\rm eff}$ a
gauge invariant functional of
the $\cN=2$ Yang-Mills multiplet.
Without loss of generality,
$\cL^{(+4)}_{\rm eff}$
can be expressed in the form
\be
 \cL^{(+4)}_{\rm eff}[\cW, \bar \cW ]
~= ~ (\cD^+)^4  \cL_{\rm eff} [\cW, \bar \cW ]~,
\ee
with $\cL_{\rm eff} [\cW, \bar \cW ]$ harmonic-independent,
and hence $\G^{(1)}$ can be rewritten as
\be
\G^{(1)} = \int {\rm d}^{12}z \; \cL_{\rm eff} [\cW, \bar \cW ]~.
\ee
More exotic quantum corrections
$$
{\rm tr}\; \Big\{ (\cD^+)^2 f(\cW)\; (\cD^+)^2 g(\cW)\Big\}
\quad + \quad {\rm c.c.}
$$
 conflict with superconformal symmetry,
although such corrections   are not forbidden by the above
conditions on $\cL^{(+4)}_{\rm eff}$.

\sect{On-shell background}

If the background gauge multiplet is on-shell,
$\cD^+ \cD^+ W=0$,  the analytic d'Alembertian does
not involve any harmonic derivative $\cD^{--}$ and
eq. (\ref{working}) takes the simpler form
\bea
\P^{(2,2)}_{\rm T}(1,2) &=&
(\cD^+_1)^4 \;\d^{12} (z_1 -z_2)\;(u^-_1 u^+_2)^2  \\
&-& (u^+_1 u^+_2)\;
\frac{1}{ {\stackrel{\frown}{\Box}}_1}
(\cD^+_1)^4
\left\{ (u^+_1 u^+_2)\;(\cD^-_1)^4
-\frac{\rm i}{2} (u^-_1 u^+_2) \;\D^{--}_1 \right\}
\;\d^{12} (z_1 -z_2)~.\non
\eea
Using this and the identities
\be
(u^-_1 u^+_2)|_{1=2} = -1~, \qquad \quad
(u^+_1 u^+_2)|_{1=2} = 0,
\ee
one immediately observes that the structure of
the effective action (\ref{starting}) drastically simplifies to
\be
\G^{(1)} = -\frac{\rm i}{2} \int_{0}^{\infty}
\frac{ {\rm d}t}{t}\; \int {\rm d} \z^{(-4)}_1\;
{\rm tr} \left(
{\rm e}^{-{\rm i} \,t\,{\stackrel{\frown}{\Box}}_1 }\;
(\cD^+_1)^4\;\d^{12} (z_1 -z_2)\right)\Big|_{z_1=z_2}~.
\label{working-on-shell}
\ee

The representation (\ref{working-on-shell}) allows us to
compute $\G^{(1)}$ in a manner almost identical to
$\cN=1$ superfield heat kernel calculations \cite{MG,PB}. This is
facilitated by the similarity of the on-shell $\cN=2$ analytic
d'Alembertian
\begin{eqnarray*}
{\stackrel{\frown}{\Box}}{}&=&
{\cal D}^m{\cal D}_m+
\frac{{\rm i}}{2}({\cal D}^{+\a}\cW){\cal D}^-_\a+\frac{{\rm i}}{2}
({\bar{\cal D}}^+_{\dot\alpha}{\bar \cW}){\bar{\cal D}}^{-{\dot\alpha}}
+ \frac{1}{2}\{{\bar \cW},\cW \}
\end{eqnarray*}
to the $\cN=1$ full superspace  d'Alembertian
\begin{eqnarray*}
{\stackrel{\frown}{\Box}}{}&=&
{\cal D}^m{\cal D}_m -
\cW^{\a}{\cal D}_\a - {\bar \cW}_{\dot\alpha}{\bar{\cal D}}^{{\dot\alpha}},
\end{eqnarray*}
as well as the similarity of the $\cN=2$ analytic delta function
$(\cD^+_1)^4\;\d^{12} (z_1 - z_2)$ with the $\cN=1$ full superspace delta
function $\d^{8} (z_1 - z_2).$
{} For simplicity, here we choose the gauge group to be
$SU(2)$ and restrict our
consideration to the Coulomb branch of the theory,
\be
[ \cW \, , \, \bar \cW] ~=~0~,
\ee
and explicitly $\cW =\hf \s_3 \,W$. On the Coulomb branch,
one can  use a derivative expansion for
the effective action. Ignoring all contributions
with vector derivatives of $W$ and $\bar W$,
direct heat kernel calculations of the effective action
(\ref{working-on-shell}) lead to
\bea
 \cL^{(+4)}_{\rm eff} &=&  (D^+)^4  \cL_{\rm eff} ~, \non \\
\cL_{\rm eff} &=&
\frac{1}{16 \pi^2}
\ln \frac{W}{\m}\, \ln \frac{{\bar W}}{\m} +
\frac{1}{8 \pi^2}
\int_0^{\infty}
{\rm d}t \,t\, {\rm e}^{-t}~
\O(t  \J \, , \, t \bar \J) ~,
\label{fin3}
\eea
where
\be
{\bar { \J}}^2 = \frac{1}{ {\bar W}^2}\,
D^4\, \ln \frac{W}{\m}~, \qquad
{\J}^2 = \frac{1}{  W^2}\,
{\bar D}^4\, \ln \frac{\bar W}{\m}~,
\label{scs}
\ee
and $\O (x,y)= \O (y,x)$ is the analytic function
related to
\be
 \o (x, y) =\o (y, x) = \frac{ \cosh x - 1}{x^2}\,
\frac{ \cosh y - 1}{y^2}\,
\frac{x^2 - y^2}{ \cosh x -  \cosh y} -\hf
\ee
by the following rule:
 if
\be
\o(x,y) = \sum_{m,n =1}^{\infty}
c_{m,n}\, x^{2m} \,y^{2n}~,
\ee
then
\be
\O(x,y) = \frac{1}{4}\sum_{m,n =1}^{\infty}
\frac{ c_{m,n} }{ m(2m+1)n(2n+1) }
\, x^{2m} \,y^{2n}~.
\ee
In (\ref{fin3}), $\m$ is a formal scale which is introduced
to make the argument of the logarithm dimensionless.
It drops out from the effective action and the superconformal
scalars (\ref{scs}).
The one-loop effective Lagrangian (\ref{fin3}) coincides with
the one found in \cite{BKT} on the basis of
$\cN=1$ heat kernel calculations and
$\cN=2$ superconformal considerations.

\sect{Off-shell background}

{}For a generic off-shell background superfield, we have to take
into account numerous terms containing factors of the classical equation of
motion. The main role of such additional terms is to combine
with the structures present in the on-shell case
to make $\cL^{(+4)}_{\rm eff}$ analytic. We illustrate
this statement using the example of the leading term
in  (\ref{fin3}), which is
quartic in spinor derivatives:
\bea
\frac{1}{(4\pi)^2}\,(D^+)^4\,
\ln \frac{W}{\m}\, \ln \frac{{\bar W}}{\m}
= \frac{1}{(16\pi)^2}\, & \Bigg\{ &
\frac{ D^+ W\, D^+ W}{W^2}\,
\frac{{\bar  D^+} {\bar W}\, {\bar  D^+} {\bar W}}
{{ \bar W}^2} \non \\
&+& \frac{ D^+ D^+ W}{W} \,
\frac{{\bar  D^+} {\bar  D^+}{\bar W}}
{\bar W} \label{F-4}\\
&-& \frac{ D^+ D^+ W}{W} \,
\frac{ {\bar  D^+} {\bar W}\, {\bar  D^+} {\bar W}}
{{\bar W}^2}
- \frac{D^+ W\, D^+ W}{W^2}
 \frac{  {\bar  D^+} {\bar  D^+}{\bar W}} {\bar W}\;\Bigg\}  ~.\non
\eea
Only the first term on the right-hand side is present in
the on-shell case. It comes from evaluating the effective action
 with only the first term in (\ref{working})
 taken into account, in which case (\ref{starting}) yields
(with the integral over the analytic subspace  omitted):
\bea
&& -{\rm i} \int_{0}^{\infty}
\frac{ {\rm d}t}{t}\;
{\rm e}^{-{\rm i} \,t\,{\stackrel{\frown}{\Box}} }\;
(\cD^+)^4\;\d^{12} (z -z')
\Big|_{z=z'}\non \\
& = &
 -{\rm i} \int_{0}^{\infty}
\frac{ {\rm d}t}{t}\; \frac{t^4}{4}\,
 D^+ W\, D^+ W {\bar  D^+} {\bar W}\, {\bar  D^+} {\bar W}\;
{\rm e}^{-{\rm i} \,t\,(W {\bar W }-{\rm i}\ve) }\;
{\rm e}^{-{\rm i} \,t\,{\Box} }\; \d^4 (x-x')\Big|_{z=z'} + \cdots \non \\
&=& \frac{1}{(16\pi)^2}\,
D^+ W\, D^+ W {\bar  D^+} {\bar W}\, {\bar  D^+} {\bar W}
\int_{0}^{\infty}
{\rm d}t\, t\; {\rm e}^{-t\,W {\bar W } }~ + \cdots,
\eea
where we have used the identity
\be
(\cD^-)^4\;(\cD^+)^4\; \d^8(\q-\q')\Big|_{\q=\q'}=1
\ee
and applied Schwinger's rotation
$t \rightarrow -{\rm i}\, t$ of the proper-time parameter.
The second term in (\ref{F-4}) originates from taking
into account the contribution in (\ref{working}) which
is proportional to $(\cD^+_1)^4\,(\cD^-_1)^4$;
here the operator $\cD^{--}$ appearing in
${\stackrel{\frown}{\Box}}$ must hit factor(s) of
$(u^+_1 u^+_2)$ in order to get a non-vanishing contribution in
the limit $u_1 =u_2$. Using ${\stackrel{\frown}{\Box}}^{-2} =
\int_0^{\infty} {\rm d}s\; s\;
{\rm e}^{-{\rm i} \,s\,{\stackrel{\frown}{\Box}} },$ one obtains
\bea
&-&\frac{1}{4} (D^+D^+W)
 \int_{0}^{\infty}
\frac{ {\rm d}t}{t}\;
\int_{0}^{\infty} {\rm d}s\; s\;
{\rm e}^{-{\rm i} \,(t+s)\,{\stackrel{\frown}{\Box}} }\;
(u^+ u'^+)(u^- u'^+) \d^{12}(z-z')
\Big|_{z=z', u=u'} \non \\
&-& \int_{0}^{\infty}
\frac{ {\rm d}t}{t}\;
\int_{0}^{\infty} {\rm d}s\;
{\rm e}^{-{\rm i} \,(t+s)\,{\stackrel{\frown}{\Box}} }\;
(u^+ u'^+)^2 \d^{12}(z-z')
\Big|_{z=z', u=u'}  \\
&= & \frac{1}{(16\pi)^2}\,
(D^+D^+W)^2  \int_{0}^{\infty}
{\rm d}t\,
\int_{0}^{\infty} {\rm d}s\;
\frac{
{\rm e}^{-(t+s)\,W {\bar W } }
}
{t+s} + \cdots \non \\
&=& \frac{1}{(16\pi)^2}\,
\frac{ D^+ D^+ W}{W} \,
\frac{{\bar  D^+} {\bar  D^+}{\bar W}}
{\bar W} ~ + \cdots.\non
\eea
{}Finally, the terms in the third line of  (\ref{F-4})
 emerge when we take into account
the structure  in (\ref{working})
involving the operator $\D^{--}$.
\vspace{5mm}

\noindent
{\bf Note added:} While preparing this paper,
we recognized how to extend the approach advocated
here to the case of
generic $\cN=2$ super Yang-Mills models.
The details will be reported in
a forthcoming publication \cite{KM}.

\vskip.5cm

\noindent
{\bf Acknowledgements:}
We are grateful to Peter Bouwknegt for hospitality
at the University of Adelaide where part of this work was done.

\end{document}